# A Nonlinear ODE System for the Unsteady Hydrodynamic Force – A New Approach

Osama A. Marzouk,

*Abstract*—We propose a reduced-order model for the instantaneous hydrodynamic force on a cylinder. The model consists of a system of two ordinary differential equations (ODEs), which can be integrated in time to yield very accurate histories of the resultant force and its direction. In contrast to several existing models, the proposed model considers the actual (total) hydrodynamic force rather than its perpendicular or parallel projection (the lift and drag), and captures the complete force rather than the oscillatory part only. We study and provide descriptions of the relationship between the model parameters, evaluated utilizing results from numerical simulations, and the Reynolds number so that the model can be used at any arbitrary value within the considered range of 100 to 500 to provide accurate representation of the force without the need to perform time-consuming simulations and solving the partial differential equations (PDEs) governing the flow field.

## I. Problem Description

WHEN a uniform flow is interrupted by an infinite-length cylinder, whose axis is perpendicular to the flow, there is a threshold of undisturbed velocity (for a certain fluid viscosity and cylinder diameter) over which a Hopf bifurcation occurs and the steady symmetric wake becomes periodic due to the existence of alternating vortices being shed in the wake at a nondimensional frequency, using the cylinder diameter and undisturbed velocity, that is known as the Strouhal number (St). Experimental and numerical studies showed that this threshold corresponds to a nondimensional velocity, using the cylinder diameter and kinematic fluid viscosity, that is known as the Reynolds number (Re) near 50 [1]–[3]. Before the bifurcation, the resultant hydrodynamic force on the cylinder, due to surface pressure and shear stresses, is parallel to the undisturbed flow. When the threshold Reynolds number is exceeded, this force becomes alternating in both the amplitude and direction, but its average has a non-zero value in the parallel direction. It is a common practice in hydrodynamics and aerodynamics to decompose this force into two orthogonal projections and represent them as a 'nondimensional' lift coefficient $C_L$ and 'nondimensional' drag coefficient $C_D$ by scaling these projections with a reference force (per unit length) that is the product of the fluid density, square of the undisturbed velocity, and the cylinder diameter.

In this study, we consider the total force directly but we still use a nondimensional representation, which we denote as the total-force coefficient ($C_T$). We denote the angle of the total force relative to the undisturbed-flow direction (being positive in the clockwise direction) by $\beta$ as illustrated in Fig. 1. Typical

e-mail: omarzouk@vt.edu.

histories of $C_T$ and $\beta$ (at Re = 300) are presented in Fig. 2. In these figures (and all subsequent ones), the time is normalized using the cylinder diameter and the undisturbed velocity. These histories require performing numerical simulation and solving the incompressible Navier-Stokes equations (continuity and momentum conservation laws), which govern the unsteady velocity and pressure fields of the constant-density, constant-temperature fluid, which is followed by a post-processing step to compute the surface pressure and shear stress at the cylinder surface and then integrating them to find the resultant force. Whereas well-established techniques and numerical schemes have been developed for this purpose [4]–[6], solving this system of nonlinear, coupled partial differential equations with high accuracy requires a lot computational resources and time. It is therefore desirable to circumvent these simulations by replacing this PDE system by a reduced ODE one that still describes accurately $C_T$ and $\beta$ but with much less computational demands and without the need of post-processing of the velocity and pressure fields. This is the main objective of our study. Whereas details about the flow field can not be obtained through the reduced-order model, the force on the cylinder is of primary importance when we are concerned with the design of a cylinder (or a pipe) and the implied fatigue problem.

A full cycle of $\beta$ corresponds to a full cycle of shedding where two contra-rotating vortices are shed in the wake. We marked four equally-spaced instants of time over approximately one cycle of $\beta$ starting from a point where $\beta$ is maximum. The corresponding distributions of the surface pressure (nondimensional) at these instants are presented in Fig. 3, and we superimposed an arrow indicating $C_T$ and its direction. We should mention here that we only show the distributions of the surface pressure and not the surface shear because the former is very small compared to the latter (except at very low Re before shedding takes place, which is not of interest here) as also indicated in Ref. [7]. However, in all the results of $C_T$ and $\beta$ we obtained through solving the Navier-Stokes equations and present in this study, we accounted for the contribution of both the pressure and shear. The average (over time) surface pressure distribution is symmetric and the average $C_T$ is not zero but the average $\beta$ is zero as indicated in Fig. 4.

For the same case of Re=300, we present in Fig 5 typical spectra of $C_T$ and $\beta$. We normalize the frequencies by the Strouhal number. As mentioned earlier, the fundamental frequency of $\beta$ is the shedding frequency. The fundamental frequency of $C_T$ is twice the fundamental frequency of $\beta$, thus is twice the Strouhal number. The total-force coefficient can be represented as a superposition of an average value (denoted as $a_{0T}$) and a harmonic function with a frequency at 2St and

  

amplitude $a_{2T}$, which is the magnitude of the $C_T$ harmonic at a frequency equal to 2St. All odd harmonics and other even harmonics of $C_T$ are negligible as indicated in Fig 5. On the other hand, all even harmonics of $\beta$ (at even multiples of St) are negligible. The magnitude of the third harmonic $a_{3\beta}$ is one order of magnitude smaller than the fundamental harmonic (whose magnitude is $a_{1\beta}$), and the higher odd harmonics are negligible. These features of $C_T$ and $\beta$ are common for the Re range we consider here (from 100 to 500), and we will utilize them when proposing our new ODE system that models $C_T$ and $\beta$.

## II. MODELING OVERVIEW AND OUR OBJECTIVES

Birkhoff and Zarantonello [8] observed that the wake of a stationary cylinder, with its continuous swinging, can be modeled by an oscillator and alluded to a linear one. Seven years later, Bishop and Hassan [9] conducted a thorough experimental study of a cylinder forced to oscillate perpendicular to a uniform flow. When the forcing frequency was close to the Strouhal number, the wake responded at the forcing frequency and not at the Strouhal number, and we speak of a wake synchronized with the motion of the cylinder. Based on this observation, Bishop and Hassan proposed that the periodic wake can be modeled by a simple oscillator. They did not specify a particular oscillator (which they called 'wake' or 'fluid' oscillator), but indicated that it is nonlinear and self-excited. Since then, several models have been proposed for the wake of a rigid cylinder [10]–[23]. Each model is a nonlinear ODE that, when integrated in time, yields the history of the lift coefficient.

Two candidate models have been commonly used so far for the lift force on a cylinder in a uniform flow: the Rayleigh oscillator [10], [11], [13], [14], [19] and the van der Pol oscillator [20]–[23]. The van der Pol oscillator for the lift coefficient ($C_L$) has the form

$$\ddot{C}_L + \omega^2\, C_L + \mu\, \dot{C}_L + c\, C_L^2\, \dot{C}_L = 0 \qquad (1)$$

and the Rayleigh oscillator has the form

$$\ddot{C}_L + \omega^2\, C_L + \mu\, \dot{C}_L + c\, \dot{C}_L^3 = 0 \qquad (2)$$

Both are self-excited self-limiting nonlinear oscillators with a cubic nonlinearity. Krenk and Nielsen [18] proposed a combination of these two oscillators, and Griffin et al. [24] and Skop and Griffin [25] even added a Duffing term ($C_L^3$). Landl [15] proposed a variant of the van der Pol oscillator with an additional quintic term ($C_L^4\, \dot{C}_L$) as an additional hydrodynamic damping, but this increased the model parameters to be determined and its complexity. It should be noted that only odd nonlinearities can be considered in the wake oscillator for $C_L$ because its spectrum mainly consists of a dominant component at St and harmonics at odd multiples of St.

It was Nayfeh et al. [20] who justified their choice of the van der Pol oscillator, rather than the Rayleigh one, to describe the lift coefficient exerted on a stationary cylinder by the wake. They reinforced that the van der Pol oscillator produces a phase of the third harmonic, at 3St, relative to the fundamental one, at St, that is equal to $90^o$ as compared to $270^o$ in the case of the Rayleigh oscillator. They found that corresponding phase obtained from numerical simulations (solving the Reynolds-averaged Navier-Stokes equations at different Re) was approximately $90^o$. Qin [22] and Marzouk et al. [23] based their choice of the same oscillator on the reasoning of Nayfeh et al. [20], whereas Facchinetti et al. [21] did not give a reason for choosing this oscillator. The wake oscillator of Facchinetti et al. [21] lacks the flexibility of describing the changes in the lift coefficient at different wake regimes (different Reynolds numbers) because the coefficient of the (negative) linear damping and (positive) nonlinear damping are equal, thus the model yields a unique limit cycle irrespective of the Reynolds number, which is a serious violation of the physics of the problem. The model also assumes a constant Strouhal number of 0.2, which also should be allowed to vary with the Reynolds number as found in experimental and numerical studies (see Appendix A). We note that the model variable in the work of Facchinetti et al. [21] is the $C_L$ scaled to a reference constant and not the $C_L$ itself. However, this does not affect the model dynamics.

Marzouk et al. [23] added a Duffing term to the van der Pol oscillator. Although Griffin et al. [24] and Skop and Griffin [25] have previously included a Duffing term in their mixed (van der Pol and Rayleigh) wake oscillator, they did not provide a reason behind including such term, which seems to be just an attempt to generalize the oscillator. In contrast, Marzouk et al. [23] explained the need for this term by the ability of the model to capture exactly the relative phase of the third harmonic of $C_L$, which would be fixed at $90^o$ without this Duffing term. They also provided expressions relating the model parameters to the $C_L$ characteristics (e.g., the frequency and magnitude of the fundamental harmonic).

Ogink and Metrikine [26] started from where Facchinetti et al. [21] ended and modified the cubic term in the van der Pol oscillator from $C_L^2\, \dot{C}_L$ to $1/(a + b\, C_L^2)\dot{C}_L$, where $0 < a < 1$ and $0 < b$ are two tuning parameters. This modification was proposed so that for increasing values of $C_L$, the influence of the nonlinearity in the modified model decreases. However, they did not discuss how $a$ and $b$ are related to the Reynolds number or the wake characteristics. They set these extra tuning parameters to constant values of $a=1/2$ and $b=1$.

Some studies [16], [20], [22], [23], [27] also considered a reduced-order model for the drag. Currie and Turnbull [16] used a Rayleigh oscillator to describe the oscillatory part only of the drag coefficient. Assuming a harmonic oscillatory drag, a relationship between the parameters of the linear and nonlinear damping was established such that the amplitude of the oscillatory drag coefficient becomes 0.2, and the model does not allow any variation of this value with the Reynolds number. This is a remarkable drawback in the model, which also specifies the individual values of these parameters based on matching the model results to an arbitrary data set in a trial-and-error fashion.

Kim and Perkins [27] considered an elastic cable suspension, and introduced a drag model in which the oscillatory part of the drag coefficient was represented as the product of a time-dependent function and a spatial function whose argument is the longitudinal coordinate (divided by the cable diameter) along the cable in the static condition. The time-dependent function was modeled by a van der Pol oscillator whose natural frequency is at twice the Strouhal number. A similar oscillator was used for the lift coefficient but the natural frequency is at the Strouhal number. Kim and Perkins used seven quadratic coupling terms in the two oscillators with seven parameters that need to be identified. These parameters were grouped in four functions, whose values were simply selected manually to fit some experimental data, which also was followed in determining other model parameters for the uncoupled part of the oscillators. Therefore, their coupled lift and drag models have many parameters to identify; there is no robust method of determining these parameters, and the average drag is not reproduced.

Nayfeh et al. [20], Qin [22], and Marzouk et al. [23] assumed the oscillatory part of the drag coefficient to be a harmonic function at twice the Strouhal number and modeled it by an algebraic function of the independently-modeled lift coefficient. Nayfeh et al. [20] used a single-term quadratic function (proportional to $C_L \, \dot{C}_L$). This freezes the phase between the drag and lift to $270^o$. Qin [22] used another single-term quadratic function (proportional to the oscillatory part of $C_L^2$) and added a linear coupling term (proportional to $C_L$). However, our simulations show that such linear coupling (which was proposed only by Qin) is not necessary. Marzouk et al. [23] extended the drag model of Nayfeh et al. [20] by combining their coupling term with the one used by Qin [22], so that the resulting two-parameter algebraic model can reproduce any value of the relative drag phase. The coefficients of these terms were analytically related to the relative drag phase.

Marzouk and Nayfeh [28] modeled the total-force coefficient and its angle, and considered a mixed van der Pol and Duffing nonlinear terms for the angle oscillator, but used an algebraic quadratic equation to relate the oscillatory part of the total-force coefficient to the evolving angle. Thus, their differential/algebraic wake model cannot resolve the average value of the total-force coefficient. In addition, they considered a single Reynolds number. The present study addresses these limitations.

The quality of a reduced-order model for the force on a cylinder due to the periodic distributions of the pressure and shear stresses on its surface as a result of the near-wake shedding process is gauged by the capability of the model to capture qualitatively and quantitatively the main characteristics of this force, which inevitably depend on the Reynolds number and not universal constants. Therefore, we propose in this study a new wake model, which consists of two coupled ODEs for the nondimensional (or coefficient $C_T$) resultant force of the instantaneous surface stresses and its angle. The model has three advantages: First, it directly models the actual hydrodynamic force rather than modeling the lift and drag, which are two 'convenient' components of this actual force. It is still possible to retrieve the lift and drag in a simple post-processing step of the resolved total force, where $C_L = C_T \sin(\beta)$ and $C_D = C_T \cos(\beta)$. Second, we provide closed-form expressions relating the parameters of the model to the main flow characteristics, which depend on the Reynolds number. Therefore, our ODE system implicitly accounts for the variations of the wake regime with the Reynolds number and does not freeze any of these characteristics (such as the Strouhal number or amplitude of the lift coefficient). The model parameters can be obtained in a schematic way which is faster and more accurate than manually matching these parameters to match a target set of data. Third, our proposed ODE system captures directly the average component of the total force (thus the average drag) and does not ignore it. This, in addition to directly modeling the actual hydrodynamic force, makes the model better than the existing ones in terms of being able to mathematically describe the physical aspects of the problem with simple equations which are much easier to solve than the full Navier-Stoked equations. We provide continuous functions of the model parameters, within the considered range of Reynolds number, so that one can use the proposed model at any arbitrary value of Reynolds number, which can then be coupled with a structural oscillator to investigate the two-degree-of-freedom fluid-structure interaction between the elastically-mounted cylinder and the flow. The modeling concept we followed here is not strictly limited to a circular cylinder. Rather, it can be extended to other geometries that exhibit a similar shedding phenomenon [29].

## III. Proposed ODE System

As indicated in the simulation results in Fig. 5, the spectrum of $\beta$ consists of a dominant component at St (with magnitude $a_{1\beta}$) and a third harmonic at 3St (with magnitude $a_{3\beta}$). Such behavior of $\beta$ suggests a self-excited oscillator such as the van der Pol or the Rayleigh oscillators. Looking at the phase by which of the component at 3St leads the one at St (which we denote by $\psi_{\beta3}$) over the range of Re under consideration, we found it to lie between $148.8^o$ and $154.5^o$. Therefore, the van der Pol oscillator (which implies a modeled $\psi_{\beta3}=90^o$) is preferred to the Rayleigh oscillator (which implies a modeled $\psi_{\beta3}=270^o$). However, we still need to adjust the van der Pol oscillator by either combining it with a Rayleigh oscillator or a Duffing one, so that the weights of the two cubic-nonlinearity terms in the combined oscillators can be adjusted so that the modeled $\psi_{\beta3}$ is precisely equal to a desired value that is known to occur at a certain Re. Whereas for the case of modeling the force on a stationary cylinder, there is no obvious preference of which oscillator to combine with the van der Pol oscillator, we examine the case of a cylinder undergoing harmonic oscillations to make this choice. Experimental [9], [30], [31] and numerical [32]–[34] studies at low and high Reynolds numbers show that the frequency-response curves of the $C_L$, $C_D$, and surface pressure are asymmetric about the St. When the oscillation amplitude is high enough, hysteresis and jumps occur, which coincide with a transition between two different wake modes (one having stronger forces than the other). The forced van der Pol, Rayleigh, or a combination exhibits symmetric frequency-response curves about the linear frequency, and two symmetrically-located hysteresis locations would be observed in that case [35], which does not match the physics of the problem. On the other hand, adding the Duffing term to the van der Pol oscillator can produce biased frequency-response curves with single location of hysteresis [36] because the backbone curve is no longer vertical (as in the case of forced combination of the van der Pol or Rayleigh oscillators). Therefore, we propose a combined (van der Pol and Duffing) oscillator for $\beta$.

It is desired to use a linear oscillator for the amplitude of the total-force coefficient $C_T$ because of its simplicity and to reduce the number of overall-model parameters. However, proper forcing is needed (unlike the free $\beta$ oscillator) in order to obtain a stable nontrivial $C_T$ response. A functional form of $\beta$ is a good candidate for this purpose because physically it is the complementary feature needed with $C_T$ to fully describe the evolution of the total force coefficient. Utilizing the two-to-one frequency relationship between $C_T$ and $\beta$, we set the linear frequency of the $C_T$ oscillator to be twice the frequency of the $\beta$ oscillator. We seek a forcing of the $C_T$ oscillator that satisfies three conditions. First, it is at a frequency of 2St (which is the correct frequency of the $C_T$). Second, it has an average value, which provides the average $C_T$. Third, it has at least two terms, so that the resultant phase by which $C_T$ leads $\beta$ is not a universal constant and can vary with the Reynolds number. Under these conditions, the simplest form of the forcing is a combination of $\beta^2$ and $\beta\,\dot{\beta}$. We note that $\beta^2$ could be replaced by $\dot{\beta}^2$, but we will use the latter, mainly for convenience as $\beta$ is a more 'visible' quantity than $\dot{\beta}$.

Based on this discussion and justifications, we propose the following system to describe the time-dependent total-force coefficient ($C_T$) and its angle, in radians, ($\beta$):

$$\ddot{\beta} + \omega^2\,\beta + \mu_1\,\dot{\beta} + c_1\,\beta^2\,\dot{\beta} + c_2\,\beta^3 = 0 \tag{3a}$$

$$\ddot{C}_T + 4\omega^2\,C_T + \mu_2\,\dot{C}_T = q_1\,\beta^2 + q_2\,\dot{\beta}\,\beta \tag{3b}$$

The proposed system in Eq. (3) has seven model parameters; namely the linear frequency $\omega$, the linear 'destabilizing' damping coefficient $\mu_1 < 0$, the cubic 'stabilizing' damping coefficient $c_1 > 0$, the Duffing-term coefficient $c_2$, the linear 'stabilizing' damping coefficient $\mu_2 > 0$, the quadratic static-static coupling coefficient $q_1$, and the quadratic static-kinematic coupling coefficient $q_2$. In the remaining part of this section, we analyze the model and derive expressions for these parameters, which analytically relate them to seven target flow characteristics, which we assume are known (e.g., from processing of a simulated or measured flow field or from interpolating the parameters from pre-determined values at different Reynolds numbers). These flow characteristics are the Strouhal number (St); the harmonics magnitudes $a_{1\beta}$, $a_{3\beta}$, $a_{2T}$; and the relative phases $\psi_{\beta3}$ and $\psi_{T2}$. We use the method of harmonic balance and find first the algebraic equations for the four parameters in Eq. (3a) and then for the three remaining parameters in Eq. (3b). In the following analysis, we do not make any assumptions about the relative magnitudes of these parameters.

A general Fourier series for the periodic $\beta$, with the dominant component and its small superharmonic component can be written as

$$\beta(t) = a_{1\beta}\cos(2\pi\,\mathrm{St}\,t + \sigma_\beta) + a_{3\beta}\cos(6\pi\,\mathrm{St}\,t + 3\sigma_\beta + \psi_{\beta3}) \tag{4}$$

Because $\beta$ has a self-excited limit cycle and no external forcing appears in Eq. (3a), $\sigma_{\beta1}$ is arbitrary. So, we can set $\sigma_{\beta1} = 0$. With this, we re-write Eq. (4) as

$$\beta(t) = a_{1\beta}\cos(2\pi\,\mathrm{St}\,t)$$
$$+ a_{3\beta}\cos(6\pi\,\mathrm{St}\,t)\cos(\psi_{\beta3})$$
$$- a_{3\beta}\sin(6\pi\,\mathrm{St}\,t)\sin(\psi_{\beta3}) \tag{5}$$

Substituting Eq. (5) into Eq. (3a), and then collecting the coefficients of $\cos(2\pi\,\mathrm{St}\,t)$, $\sin(2\pi\,\mathrm{St}\,t)$, $\cos(6\pi\,\mathrm{St}\,t)$, and $\sin(6\pi\,\mathrm{St}\,t)$ results in the following system of algebraic equations:

$$\chi_1\,\omega^2 \quad + \qquad\quad \chi_3\,c_1 \; + \chi_4\,c_2 \;\; = \chi_5 \tag{6a}$$

$$\qquad\quad \delta_2\,\mu_1 \; + \delta_3\,c_1 \; + \delta_4\,c_2 \;\; = 0 \tag{6b}$$

$$\epsilon_1\,\omega^2 \; + \epsilon_2\,\mu_1 \; + \epsilon_3\,c_1 \; + \epsilon_4\,c_2 \;\; = \epsilon_5 \tag{6c}$$

$$\gamma_1\,\omega^2 \; + \gamma_2\,\mu_1 \; + \gamma_3\,c_1 \; + \gamma_4\,c_2 \;\; = \gamma_5 \tag{6d}$$

where

$$\chi_1 = a_{1\beta}; \;\; \chi_3 = \Lambda\,a_{1\beta}^2\,\eta$$
$$\chi_4 = 0.75a_{1\beta}^3 + 0.75a_{1\beta}^2\,\xi + 1.5a_{1\beta}\,\xi^2 + 1.5a_{1\beta}\,\eta^2$$
$$\chi_5 = 16\Lambda^2\,a_{1\beta}$$
$$\delta_2 = 4\Lambda\,a_{1\beta}; \;\; \delta_3 = -\Lambda\,a_{1\beta}^3 - \Lambda\,a_{1\beta}^2\,\xi - 2\Lambda\,a_{1\beta}\,\xi^2 - 2\Lambda\,a_{1\beta}\,\eta^2$$
$$\delta_4 = 0.75a_{1\beta}^2\,\eta; \;\; \epsilon_1 = \xi$$
$$\epsilon_2 = -12\Lambda\,\eta; \;\; \epsilon_3 = 3\Lambda\,\eta^3 + 6\Lambda\,a_{1\beta}^2\,\eta + 3\Lambda\,\xi^2\,\eta \tag{7}$$
$$\epsilon_4 = 0.25a_{1\beta}^3 + 1.5a_{1\beta}^2\,\xi + 0.75\xi^3 + 0.75\eta^2\,\xi; \;\; \epsilon_5 = 144\Lambda^2\,\xi$$
$$\gamma_1 = \eta; \;\; \gamma_2 = 12\Lambda\,\xi$$
$$\gamma_3 = -\Lambda\,a_{1\beta}^3 - 6\Lambda\,a_{1\beta}^2\,\xi - 3\Lambda\,\xi^3 - 3\Lambda\,\eta^2\,\xi$$
$$\gamma_4 = 0.75\eta^3 + 1.5a_{1\beta}^2\,\eta + 0.75\xi^2\,\eta; \;\; \gamma_5 = 144\Lambda^2\,\eta$$

with

$$\Lambda = \pi\,\mathrm{St}/2; \;\; \xi = a_{3\beta}\cos(\psi_{\beta3}); \;\; \eta = -a_{3\beta}\sin(\psi_{\beta3}) \tag{8}$$

Solving Eq. (6) yields the four model parameters $\omega$, $\mu_1$, $c_1$, and $c_2$.

The time-dependent total-force coefficient can be approximated as

$$C_T(t) = a_{0T} + a_{2T}\cos(4\pi\,\mathrm{St}\,t + \psi_{T2}) \tag{9}$$

which is expanded to

$$C_T(t) = a_{0T}$$
$$+ a_{2T}\cos(\psi_{T2})\cos(4\pi\,\mathrm{St}\,t)$$
$$- a_{2T}\sin(\psi_{T2})\sin(4\pi\,\mathrm{St}\,t) \tag{10}$$

We will neglect the influence of the superharmonic component of $\beta$ on $C_T$ based on our finding that $a_{3\beta}$ is one order of magnitude smaller than $a_{1\beta}$ and the linearity of the homogenous part of Eq. (3b). We substitute $\beta = a_{1\beta}\cos(2\pi\,\mathrm{St}\,t)$ into Eq. (3b) and solve for the particular solution of the forced

system as

$$C_T(t) = \frac{q_1\,a_{1\beta}^2}{8\omega^2} \tag{11a}$$
$$+ (q_1\,\tau_1 + 2\pi\,St\,q_2\,\tau_2)\cos(4\pi\,St\,t)$$
$$+ (q_1\,\tau_2 - 2\pi\,St\,q_2\,\tau_1)\sin(4\pi\,St\,t)$$

with

$$\tau_1 = \frac{a_{1\beta}^2}{2}\ \frac{4\omega^2 - 64\Lambda^2}{(4\omega^2 - 64\Lambda^2)^2 + (8\Lambda\,\mu_2)^2} \tag{11b}$$

$$\tau_2 = \frac{a_{1\beta}^2}{2}\ \frac{8\Lambda\,\mu_2}{(4\omega^2 - 64\Lambda^2)^2 + (8\Lambda\,\mu_2)^2} \tag{11c}$$

The equivalence between Eq. (10) and Eq. (11) yields the required algebraic equations for the three remaining model parameters ($\mu_2$, $q_1$, and $q_2$) as

$$q_1 = 8\,\frac{\omega^2}{a_{1\beta}^2}\,a_{0T} \tag{12a}$$

$$(2q_2\,\Lambda\,a_{1\beta}^2)^2 + (q_1\,a_{1\beta}^2/2)^2$$
$$= a_{2T}^2\ \left[(4\omega^2 - 64\Lambda^2)^2 + (8\Lambda\,\mu_2)^2\right] \tag{12b}$$

$$\arctan\left(\frac{-8\Lambda\,\mu_2}{4\omega^2 - 64\Lambda^2}\right) - \arctan\left(\frac{-4\Lambda\,\mu_2}{q_1}\right) = \psi_{T2} \tag{12c}$$

Equation (12a) can be solved directly for $q_1$, then Eqs. (12b) and (12b) are solved simultaneously for $\mu_2$ and $q_2$.

## IV. RESULTS

We performed numerical simulations and solved the Navier-Stokes equations over a range of the Re from 100 to 500, with increments of 50 except near Re=100, where the changes in the flow characteristics are fast and thus the increment is reduced to 25. We chose this range of the Re because the near-wake shedding is coherent enough to result in a regular periodic hydrodynamic force, which can be precisely modeled by a reduced-order system. At higher Re, this force exhibits fast modulations and irregularity due to increasing disorder in the wake and the intensified turbulence, thereby reflecting adversely on the feasibility of modeling it quantitatively in the time and spectral domains. In addition, this range is of special importance in this problem because it exhibits rapid changes in the flow features, such as the shedding frequency and the amplitude of $\beta$ and $C_L$. Near the end of this range and beyond it, these features become slowly-varying functions of the Re. For example, the St Changes from 0.1654 to 0.21064 as the Re changes from 100 to 300. This is larger than the change in the St for a Re range from 300 to 300,000 [3], [37], [38]. We use the artificial compressibility method and a body-fitted nonuniform grid with 43,200 grid points and a fixed nondimensional time step of 0.02. For Re<300, we solve the full Navier-Stokes directly; whereas for Re≥300, we solve the Reynolds-averaged Navier Stokes equations (RANS) in order to account for the developing turbulence in the wake. We model the turbulence effects by the Spalart-Allmaras model [39]. More details about the solution scheme can be found in Refs. [6], [23], [34], [40], [41]. In Appendix A, we compiled reported values of the St, amplitude and root mean square of $C_L$, and the average $C_D$ form experimental and other computational studies [7], [42]–[57] for six values of the Re and compared them with those obtained from our simulations in Tables I-III, respectively. Our results are in agreement with those reported from the other studies, which validates the numerical simulations we performed.

For each simulation, we calculated the surface pressure and shear stresses and integrated them (over the cylinder surface) to obtain the time-dependent $C_L$, $C_D$, $C_T$, and $\beta$. Before we examine the performance of the proposed ODE system, we examine the relationships between the Re and the flow characteristics, to be used to determine the model parameters as described before, in the Re range we considered. The relationship of the St (or the nondimensional shedding frequency) with the Re is presented in Fig. 6. The corresponding relationships of the $\beta$ characteristics ($a_{1\beta}$, $a_{3\beta}$, $\psi_{\beta3}$) are presented in Fig. 7, and the relationships of the $C_T$ characteristics ($a_{0T}$, $a_{2T}$, $\psi_{T2}$) are presented in Fig. 8. Whereas the St, $a_{1\beta}$, $a_{3\beta}$, and $a_{2T}$ always increase with the Re (first rapidly then slowly); $\psi_{T2}$, $a_{0T}$, and $\psi_{T2}$ do not follow a similar trend. Rather, they first increase and reach a maximum value before decreasing as the Re continues to increase.

We examined the proposed ODE system (as a reduced-order model to describe the dynamics of $C_T$ and $\beta$) for all the simulation cases we performed and found excellent performance where the modeled $C_T(t)$ and $\beta(t)$ from the ODE system match well those from the simulations. To illustrate this, we present comparisons between the modeled and simulated model variables at the lower and upper bound of the Re range we considered (i.e., Re=100 and 500) in Figs. 9 and 10, respectively. The modeled and simulated $\beta$ are almost identical for slow shedding (e.g., Re=100) or fast shedding (e.g., Re=500). As the shedding frequency increases, slight deviation occurs in $C_T(t)$, but the model performance is still very satisfactory. This can be explained by the increased $a_{3\beta}$, whose influence on $\beta$ was neglected.

The relationships of the seven model parameters with the Re are presented in Fig. 11. We adjusted $mu_2$ for improved matching of $C_T$, and the figure reflects the values we used. The linear frequency $\omega$ increases rapidly first with the Re before it reaches a maximum at Re=400 and then becomes nearly constant. As the Re (thus the shedding intensity and frequency) increases, the destabilizing and nonlinear damping coefficients increase. The Duffing-term coefficient is always negative (softening-type stiffness). Its absolute value increases with the Re and reaches a maximum at Re=350. The damping and two coupling coefficients of $C_T$ follow the same behavior as the Re is varied. Near Re=100, they are very large and undergo a steep decrease up to Re=150, followed by an oscillating behavior. In addition to presenting the values of these parameters at the selected Re values of the simulations, we provide the high-order polynomial functions (the R-square ranges from 0.9927 to 0.9996) for each parameter in Appendix B. The dependent variable is Re/100. As mentioned before, this provides a database of the model parameters so that the model can be fully constructed at any arbitrary value of the Re within the considered range.

## V. Conclusions

As a new approach for modeling the force on a stationary circular cylinder, we proposed a two-degree-of-freedom ODE system, having a self-excited oscillator with softening-type cubic nonlinearity for the angle of the total (resultant) hydrodynamic force and a coupled forced oscillator for the force coefficient. In addition to its accuracy, the proposed reduced-order model has several advantages over existing wake oscillators in terms of directly describing the real problem and reproduce its physical features, such as reproducing the average value of the force and describing the actual exerted force rather than one (or both) of its projections. We provided expressions for the model parameters and evaluated them over a range of Reynolds number near the wake bifurcation and onset of shedding, where rapid changes occur in the wake and the shedding frequency. The proposed model circumvents solving the PDE system governing the flow field and the post-processing needed to compute the evolution of the hydrodynamic force, provided that we are not interested in the flow field itself but only in the loads on the cylinder.

The model is not a mere fitting to sets of data, it gives insight about the relationship between the nonlinear or damping effects and the Re. By generating continuous functions of the model parameters over a range of Re, as we did here, one can use the model as accurate wake oscillator that governs the evolution of the force on the cylinder, and which can be coupled to a structural oscillator to study the fluid-structure interaction analytically. Whereas the focus here was on a circular cylinder, the presented approach can be extended to a wide range of body shapes where shedding in the wake takes place. However, a new set of model parameters (or database) is needed for each case.

APPENDIX A

COMPILATION OF FLOW CHARACTERISTICS AT DIFFERENT Re

TABLE I
STROUHAL NUMBER FROM OUR SIMULATIONS AND FROM OTHER STUDIES (ALL SIMULATIONS ARE TWO-DIMENSIONAL).

| Re | Source | Comments | St |
|----|--------|----------|-----|
| | Roshko (1953) | measurements | 0.168 |
| | Jordan and Fromm (1972) | vorticity-stream function formulation | 0.16 |
| | Braza et al. (1986) | predictor-corrector pressure method | 0.16 |
| 100 | Williamson (1989) | measurements | 0.164 |
| | Stansby and Slaouti (1993) | random-vortex method | 0.166 |
| | Shiels et al. (2001) | viscous vortex method | 0.167 |
| | Our simulations | artificial compressibility method | 0.1654 |
| | Roshko (1953) | measurements | 0.175-0.187 |
| 150 | Zhang et al. (1995) | marker-and-cell method | 0.191 |
| | Our simulations | artificial compressibility method | 0.18446 |
| | Roshko (1953) | measurements | 0.18-0.19 |
| 200 | Zhou et al. (1999) | vortex-in-cell | 0.1992 |
| | Wang et al. (2008) | RANS with SST $K - \omega$ turbulence | 0.19 |
| | Our simulations | artificial compressibility method | 0.19635 |
| | Roshko (1953) | measurements | 0.204 |
| 300 | Zhang et al. (1995) | marker-and-cell method | 0.217 |
| | Our simulations | RANS with SA turbulence | 0.21064 |
| | Roshko (1953) | measurements | 0.204 |
| 400 | Kaiktsis et al. (2007) | spectral element method | 0.22017 |
| | Our simulations | RANS with SA turbulence | 0.21432 |
| | Roshko (1953) | measurements | 0.206-0.212 |
| 500 | Blackburn and Henderson (1999) | spectral element method | 0.228 |
| | Our simulations | RANS with SA turbulence | 0.21475 |

TABLE II
AMPLITUDE AND ROOT MEAN SQUARE OF THE LIFT COEFFICIENT FROM OUR SIMULATIONS AND FROM OTHER STUDIES (ALL SIMULATIONS ARE TWO-DIMENSIONAL).

| Re | Source | Comments | Amp. of $C_L$ | RMS of $C_L$ |
|----|--------|----------|---------------|--------------|
| | Braza et al. (1986) | predictor-corrector pressure method | 0.28 | – |
| | Stansby and Slaouti (1993) | random-vortex method | 0.35 | 0.248 |
| 100 | Zhou et al. (1999) | vortex-in-cell | – | 0.219 |
| | Shiels et al. (2001) | viscous vortex method | 0.3 | – |
| | Our simulations | artificial compressibility method | 0.3249 | 0.22967 |
| | Zhang et al. (1995) | marker-and-cell method | – | 0.41 |
| 150 | Ravoux et al. (2003) | embedding method | – | 0.31 |
| | Norberg (2003) | empirical function | 0.1831 | – |
| | Our simulations | artificial compressibility method | 0.51706 | 0.36536 |
| | Braza et al. (1986) | predictor-corrector pressure method | 0.75 | – |
| 200 | Zheng and Zhang (2008) | immersed-boundary method | 0.65 | – |
| | Our simulations | artificial compressibility method | 0.67902 | 0.47981 |
| 300 | Zhang et al. (1995) | marker-and-cell method | – | 0.7 |
| | Our simulations | RANS with SA turbulence | 0.91416 | 0.64695 |
| 400 | Kaiktsis et al. (2007) | spectral element method | 0.98 | 0.76 |
| | Our simulations | RANS with SA turbulence | 1.00795 | 0.71338 |
| | Blackburn and Henderson (1999) | spectral element method | 1.2 | – |
| 500 | Ravoux et al. (2003) | embedding method | – | 0.64 |
| | Our simulations | RANS with SA turbulence | 1.00393 | 0.70875 |

TABLE III
AVERAGE OF THE DRAG COEFFICIENT FROM OUR SIMULATIONS AND FROM OTHER STUDIES (ALL SIMULATIONS ARE TWO-DIMENSIONAL).

| Re | Source | Comments | Average of $C_D$ |
|---|---|---|---|
| | Clift et al. (1978) | empirical function | 1.24418 |
| | Braza et al. (1986) | predictor-corrector pressure method | 1.29 |
| 100 | Stansby and Slaouti (1993) | random-vortex method | 1.317 |
| | Henderson (1995) | spectral element method | 1.35 |
| | Shiels et al. (2001) | viscous vortex method | 1.33 |
| | Our simulations | artificial compressibility method | 1.33804 |
| 150 | Zhang et al. (1995) | marker-and-cell method | 1.41 |
| | Our simulations | artificial compressibility method | 1.32471 |
| | Braza et al. (1986) | predictor-corrector pressure method | 1.3 |
| | Henderson (1995) | spectral element method | 1.35 |
| 200 | Zhou et al. (1999) | vortex-in-cell | 1.32 |
| | Zheng and Zhang (2008) | immersed-boundary method | 1.35 |
| | Our simulations | artificial compressibility method | 1.33779 |
| | Wieselsberger (1921) | measurements | 1.2 |
| 300 | Henderson (1995) | spectral element method | 1.4 |
| | Our simulations | RANS with SA turbulence | 1.37561 |
| | Jordan and Fromm (1972) | vorticity-stream function formulation | 1.23 |
| 400 | Kaiktsis et al. (2007) | spectral element method | 1.42 |
| | Our simulations | RANS with SA turbulence | 1.37929 |
| 500 | Blackburn and Henderson (1999) | spectral element method | 1.46 |
| | Our simulations | RANS with SA turbulence | 1.35685 |

# APPENDIX B
## FITTING FUNCTIONS FOR THE MODEL PARAMETERS

$$\omega = 0.8122 + 0.0512\,\Re - 0.3553\,\Re^2 - 0.1018\,\Re^3 + 0.0079\,\Re^4 \ (R^2 = 0.9995)$$

$$\mu_1 = 0.0526 - 0.0407\,\Re - 0.0508\,\Re^2 + 0.0073\,\Re^3 \ (R^2 = 0.9996)$$

$$c_1 = 1.6255 + 0.618\,\Re + 0.2225\,\Re^2 - 0.0341\,\Re^3 \ (R^2 = 0.9986)$$

$$c_2 = -2.1017 - 0.4127\,\Re - 2.144\,\Re^2 + 0.6189\,\Re^3 - 0.0463\,\Re^4 \ (R^2 = 0.9979)$$

$$\mu_2 = 3497.6 - 6759.5\,\Re + 5749\,\Re^2 - 2560\,\Re^3 + 629.53\,\Re^4 - 80.887\,\Re^5 + 4.2311\,\Re^6 \ (R^2 = 0.9978)$$

$$q_1 = 1299.2 - 2285.3\,\Re + 1845.7\,\Re^2 - 786.91\,\Re^3 + 187.22\,\Re^4 - 23.505\,\Re^5 + 1.2112\,\Re^6 \ (R^2 = 0.9987)$$

$$q_2 = 5957.6 - 11582\,\Re + 10030\,\Re^2 - 4543.9\,\Re^3 + 1144.6\,\Re^4 - 151.43\,\Re^5 + 8.1669\,\Re^6 \ (R^2 = 0.9927)$$

where $1 < \Re \equiv \mathrm{Re}/100 < 5$.

## Acknowledgment

The author would like to thank the Advanced Supercomputing (NAS) Division at NASA Ames Research Center. The author acknowledges the valuable advice and ideas of Professor Ali H. Nayfeh in the Department of Engineering Science and Mechanics at Virginia Polytechnic Institute and State University.

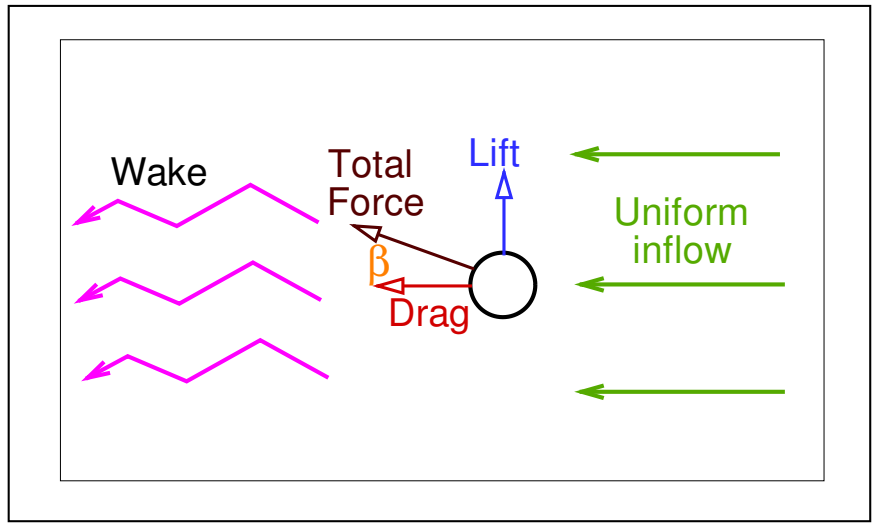

Fig. 1.   A sketch of the problem.

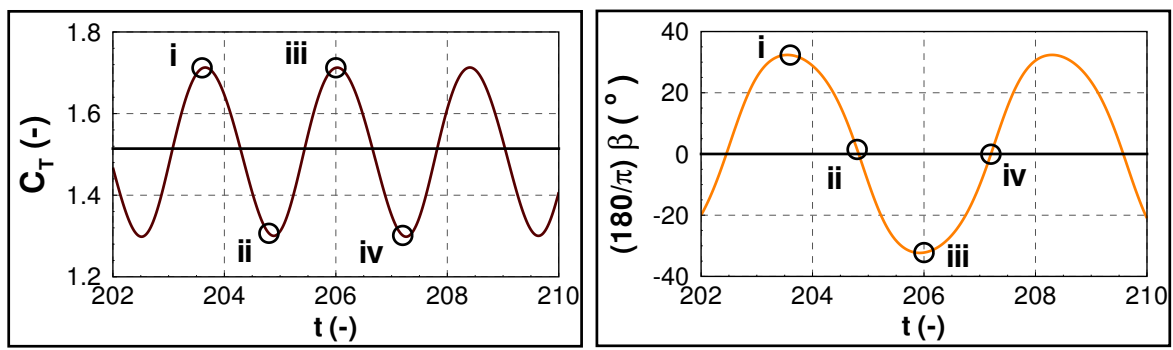

Fig. 2.   Typical histories of the total-force coefficient and its angle (at a Reynolds number of 300).

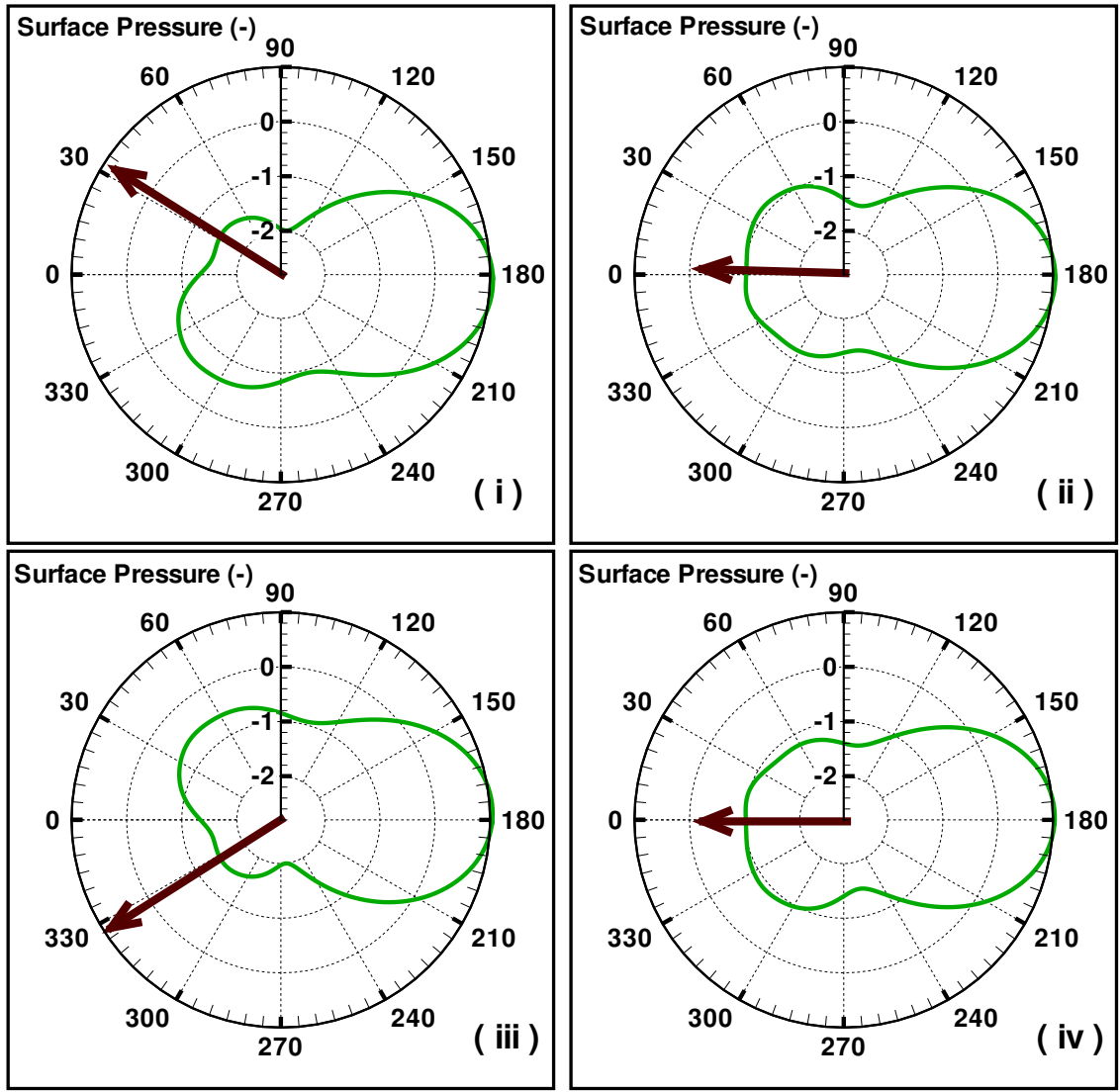

Fig. 3.   A series of the surface pressure over one shedding cycle at the four instants of time marked in Fig. 2. The undisturbed flow is encountered from the left. The total force is indicated at each plot (it has a different scale than the pressure).

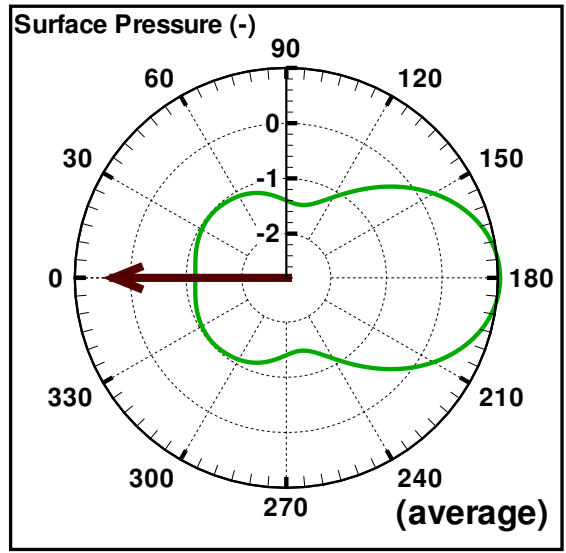

Fig. 4.  Average surface pressure corresponding to the case in Fig. 2. The undisturbed flow is encountered from the left. The average total force is indicated (it has a different scale than the pressure)

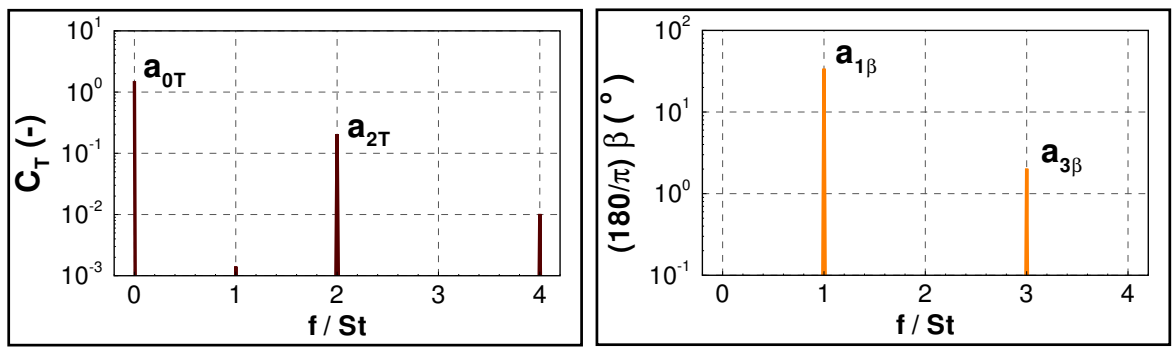

Fig. 5.  Spectra of the total-force coefficient and its angle for the case in Fig. 2.

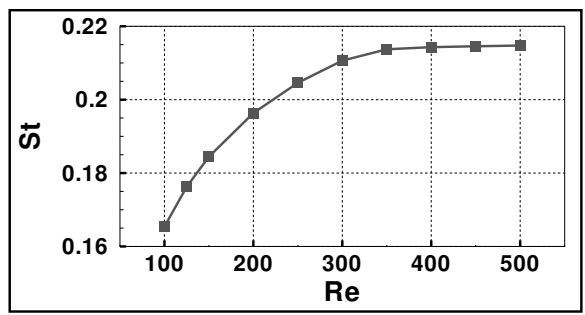

Fig. 6.  The relationship between the Strouhal number and the Reynolds number.

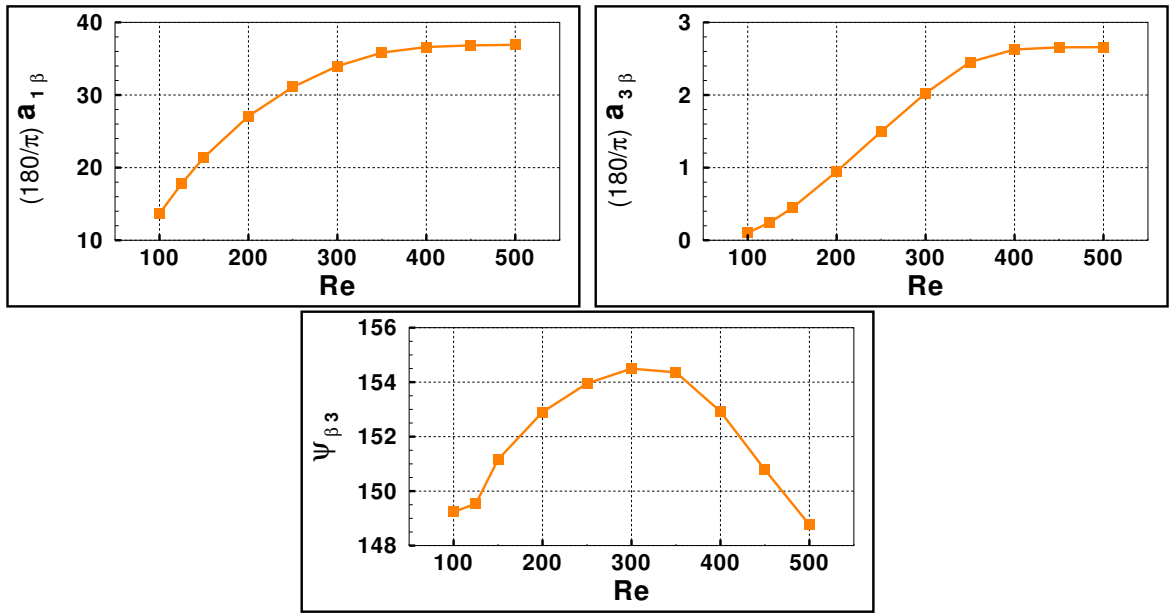

Fig. 7.   The relationships between the spectral properties of the total-force angle and the Reynolds number.

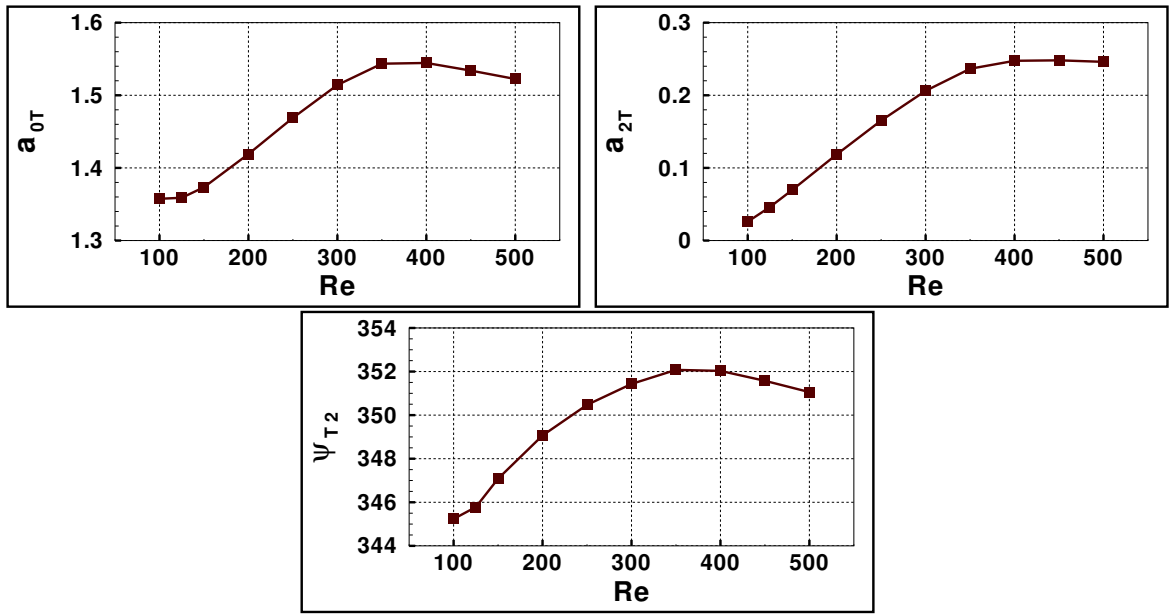

Fig. 8.   The relationships between the spectral properties of the total-force coefficient and the Reynolds number.

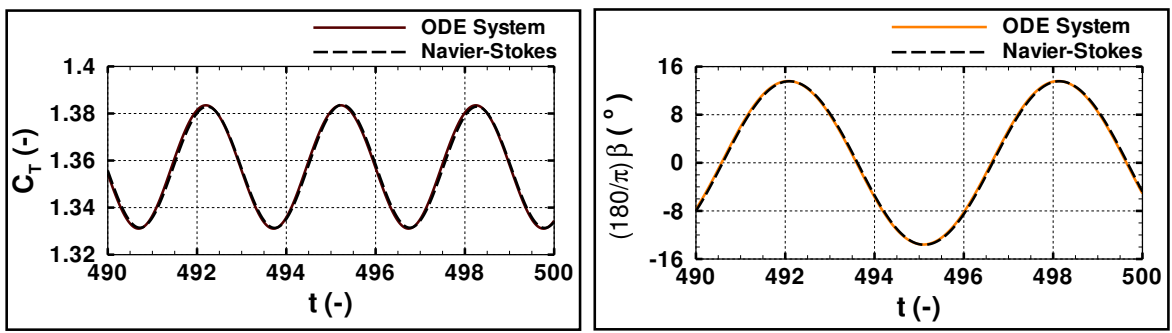

Fig. 9.   Comparison of the total-force coefficient and its angle, obtained from the proposed ODE system and from solving the Navier-Stokes equations at a Reynolds number of 100.

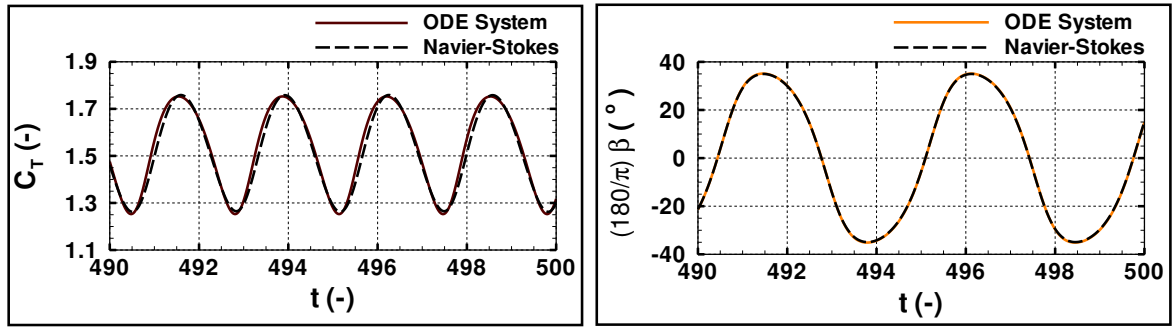

Fig. 10. Comparison of the total-force coefficient and its angle, obtained from the proposed ODE system and from solving the Reynolds-averaged Navier-Stokes equations at a Reynolds number of 500.

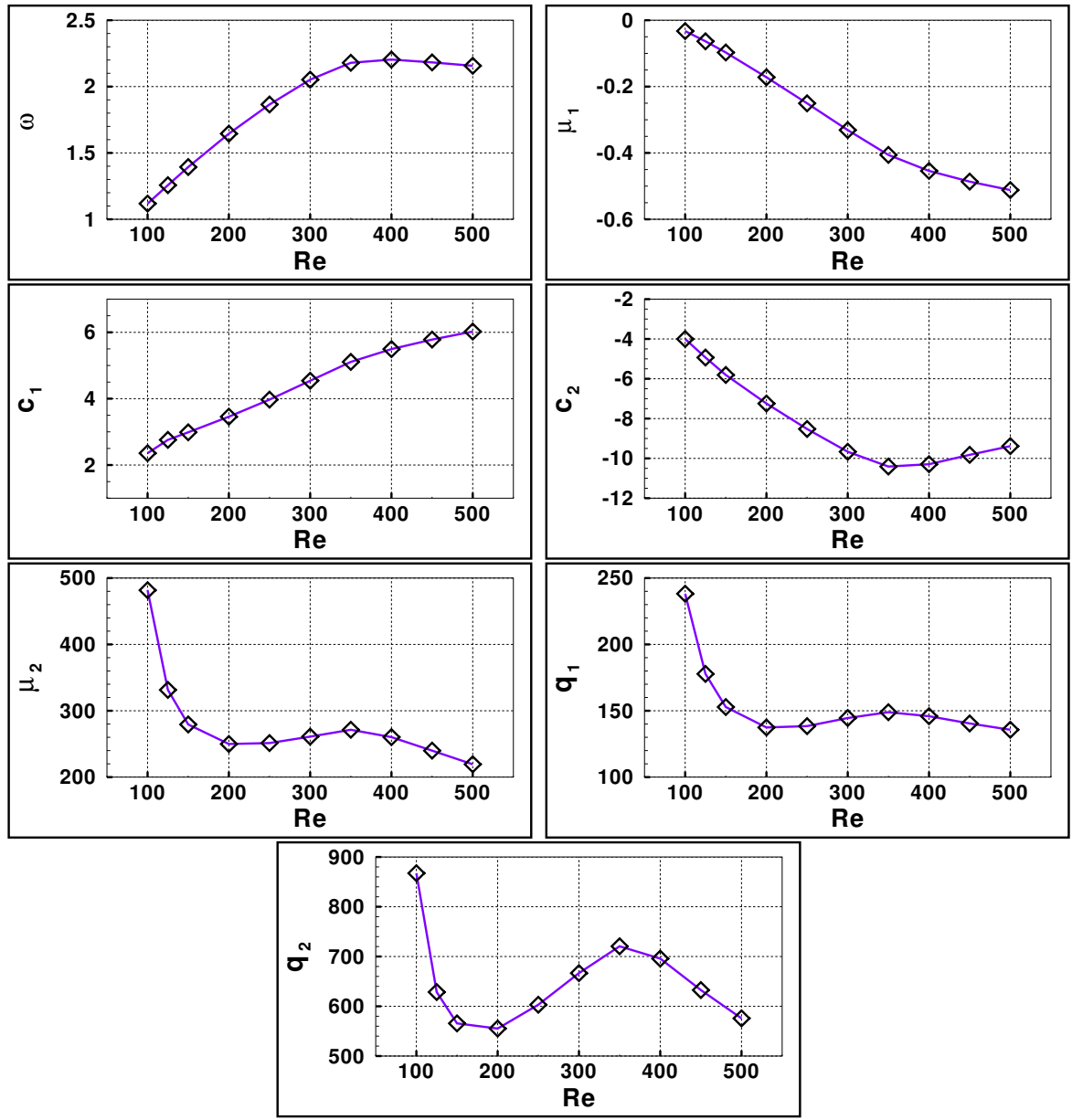

Fig. 11. The relationships between the model parameters and the Reynolds number.